\documentclass[twocolumn,aps,pre,showpacs,preprintnumbers]{revtex4}
\usepackage[dvipdf]{graphicx}
\usepackage{epstopdf}
\usepackage{epsfig} 
\usepackage{hyperref}
\usepackage{graphics}
\usepackage{color}      
\usepackage{dcolumn}

\begin{document}

\title{  HIV viral load and the efficacy  of antiviral drug     } 
\author{ Stephanie Yunfei Zhang and  Mesfin Asfaw  Taye}
\affiliation {West Los Angeles College, Science Division \\9000  Overland Ave, Culver City, CA 90230, USA}

\begin{abstract}

Developing  antiviral drugs is an exigent task  since   viruses mutate to overcome the effect of antiviral drugs. As a result,   the efficacy   of most antiviral  drugs  is  short-lived.   To include this effect, we modify the Neumann  and Dahari  model.   Considering  the  fact that the efficacy  of the antiviral drug  varies  in  time, the differential equations introduced in  the previous model systems are rewritten to study the  correlation between the viral load and   antiviral drug. The  effect of antiviral  drug that either prevents infection or  stops  the production of a virus is explored.  First,  the  efficacy  of the drug is considered  to   decreases  monotonously    as time progresses.  In this case, our result depicts that  when the efficacy  of the  drug is low, the viral load decreases  and  increases back  in time    revealing the effect of the antiviral  drugs is  short-lived.  On the other hand,  for the antiviral drug  with high efficacy, the viral load, as well as the number of infected cells, monotonously   decreases while  the number of uninfected cells   increases. The dependence of   the critical drug  efficacy   on time is also  explored. Moreover,  the correlation between  viral load, the antiviral drug,  and  CTL  response  is  also explored. In this case, not only the  dependence for the basic  reproduction ratio on the  model parameters is   explored but also we   analyze     the critical drug  efficacy   as a function of time.  We show that the term related to the  basic reproduction ratio  increases when  the CTL  response  step up.    A simple analytically solvable  mathematical model  is also presented to analyze the correlation between  viral load  and antiviral  drugs.   

\end{abstract}
\pacs{Valid PACS appear here}
\maketitle

 \section{Introduction} 
Viruses are tiny particles that occupy the world and have  property between living and non-living things. As they are not capable of reproducing, they rely on the host cells to replicate themselves. To gain access,  the virus first binds and intrudes into the host cells.  Once the virus is inside the cell, it releases its  genetic materials into the host cells.  It then  starts manipulating the cell to multiply its viral genome.  Once the viral protein is produced and assembled,  the new virus leaves the cell in search of other host cells.   Some viruses can also stay in the host cells  for a long  time as a latent or chronic state. The genetic information of a virus is stored  either in form of RNA or DNA.   Depending on the type of  virus, the  host cell type  also varies.  For instance,  the Human Immunodeficiency Virus  (HIV) (see Fig. 1 \cite{muu1})  directly affects Lymphocytes.  Lymphocytes can be categorized into two main categories: the B and T cells.  The B cells directly kill  the virus by producing a specific   antibody. T cells on the other hand  can be categorized as  killer cells (CD 8)  and helper cells (CD 4).  Contrary  to CD 8,  CD 4  gives only warning so that the cells  such as CD 8 and B cells  can directly  kill the virus \cite{mes5, muu2}. Although  most of these viruses contain DNA as genetic material,  retroviruses such as HIV   store their  genetic materials as RNA. These viruses translate their RNA into DNA using an enzyme called reverse transcriptase  during their life cycle. In the case of HIV,  once  HIV infects the patient, a higher  viral load  follows   for the first few  weeks, and then its replication  becomes steady  for several years. As a result,  the CD 4 (which is the host cell  for  HIV) decreases considerably.  When the CD 4 cells are below  a certain threshold, the patient  develops  AIDS.

\begin{figure}[ht]
\centering
{
    \includegraphics[width=6cm]{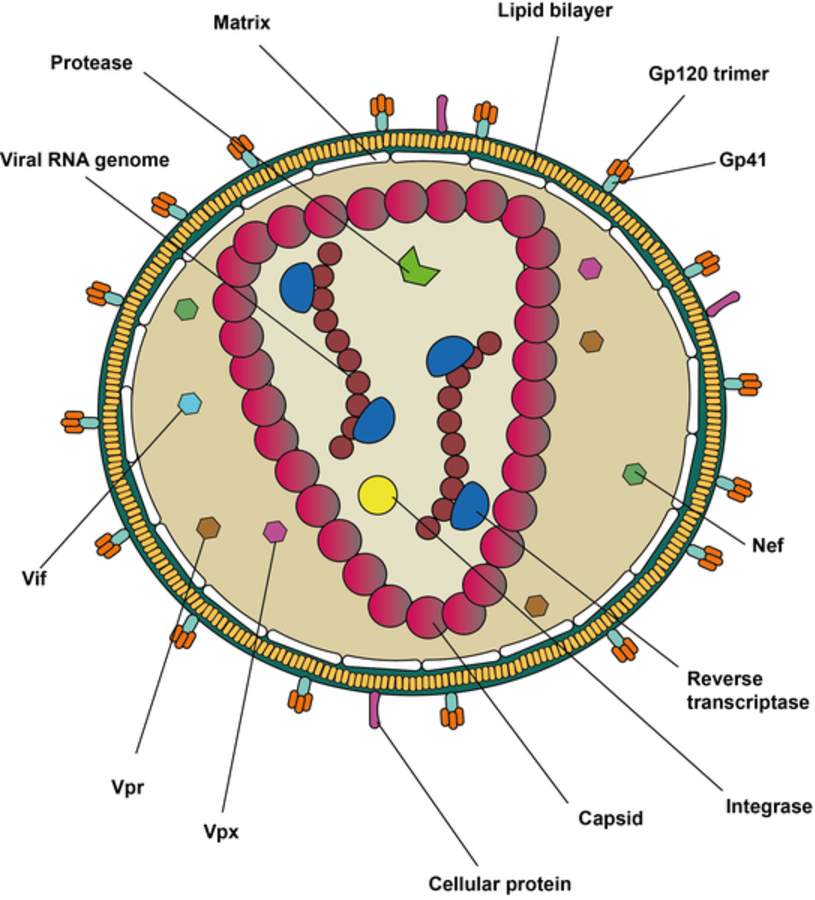}
}
\caption{ (Color online)  Schematic diagram for HIV virion \cite{muu1}.} 
\label{fig:sub} 
\end{figure}

To tackle the spread of virulent viruses such as HIV, discovering  potent antiviral drugs is vital. However,     developing  antiviral   drugs is an exigent task  since   viruses mutate to overcome the effect of antiviral drugs because of this only a few antiviral drugs are currently available. Most of these drugs are  developed to cure  HIV and herpes virus.  The fact that viruses are obligate parasites of the cells makes drug  discovery complicated since the  drug’s adverse effects directly affect the host cells.  Many medically important   viruses  are also   virulent and hence they cannot be propagated or tested via animal models.  This in turn   forces  the drug discovery to be  lengthy.    Moreover, unlike  other antimicrobial drugs, antiviral drugs  have to be 100 percent potent to completely avoid drug resistance.   In other words, if the drug partially inhibits the replication of the virus,    through time, the   number of resistant  viruses  will dominate the  cell culture.  All of the above factors   significantly hinder drug discovery.  Furthermore,  even a potent antiviral drug  does  not guarantee a cure if the acute infection is already  established.

 Understanding the dynamics  of the virus in vivo or vitro  is crucial since   viral  diseases  are  the main  global health concern.  For instance, 
recent  outbreaks of viral   diseases  such as COVID-19 not only  cost trillion dollars  but also killed more than 217,721  people in the  USA alone. To control such a global pandemic, developing an effective therapeutic strategy is vital. Particularly, in the case of virulent viruses, mathematical modeling along with the antiviral drug  helps   to  understand the dynamics of the virus in vivo \cite{mes2}. The  pioneering  mathematical models   on the Human Immunodeficiency virus  depicted in the works  \cite{mes1,mes3,mes4, mu1,mu2,mu3,mu4,mu5,mu6,mu7} shed  light regarding the host-virus  correlation.
 Latter  these model systems are modified  by Neumann $et.$ $al.$ \cite{mes1,mu8}  to study  the dynamics of HCV  during treatment. To study  the dependence  of  uninfected cells, infected cells, and virus load on model parameters, Neumann proposed three differential equations. More recently, to explore  the observed  HCV RNA profile during treatment,  Dahari $et.$ $al.$ \cite{mes1,mu9}  extended  the original  Neumann model. Their findings disclose that  critical drug  efficacy  plays  a critical role. When  the efficacy is greater  than  the critical  value, the HCV  will be cleared. On the contrary,  when the efficacy of the drug is below the critical threshold,  the   virus keeps infecting the target cells.

As discussed before, the effect  of antiviral  drugs is short-lived since the virus  mutates  during  the course of treatment.  To include this effect, we modify Neumanny and Dahariy  models. Considering  the  fact  that the efficacy  of the antiviral drug  decreases  in  time, we rewrite  the   three differential equations introduced in  the previous model systems.   The  mathematical model presented in this  work analyzes the  effect of an antiviral  drug that either prevents infection ($e_{k}$) or  stops  the production of virus ($e_{p})$.  First,  we consider a case where  the  efficacy  of the drug  decreases to zero  as time progresses and   we  then discuss  the  case where  the  efficacy  of the  drug decreases to a constant value as   time evolves  maintaining  the relation   $e_P=e_{p}'(1+e^{-rt})/m$  and  $e_r=e_{r}'(1+e^{-rt})/m$. Here  $r$,  $e_{k}'$ and  $e_{p}'$ measure  the ability of  antiviral drug  to overcome drug  resistance. When  $r$ tends to increase,  the efficacy of the drug  decreases.  The results obtained in this work depict that  for large $r$, the viral load decreases  and  increases back  as the antiviral drug  is  administered showing the effect of antiviral  drugs is short-lived.  On the other hand, for small $r$, the viral load, as well as the number of infected cells monotonously   decreases while the host cell   increases. The dependence of   the critical drug  efficacy   on time is also  explored.

The correlation between  viral load, antiviral therapy, and cytotoxic lymphocyte immune response (CTL)  is  also explored. Not only the  dependence for the basic  reproduction ratio on the  model parameters  is  explored but also we   find    the critical drug  efficacy   as a function of time.  The basic reproduction ratio  increases when  the CTL  response  decline.  When the viral load  inclines, the CTL response  step up.   We also  present a simple analytically solvable  mathematical model to address  the correlation between drug resistance and antiviral  drugs.   

The rest of the   paper is organized as follows: in Section II,  we explore the  correlation between antiviral treatment and viral load. In Section III the relation  between  viral load, antiviral therapy, and  the CTL immune response is  examined. A simple analytically solvable  mathematical model that  addresses  the correlation  between drug resistance and viral load  is presented in section IV.  Section V deals with summary
and conclusion.

\section{ The relation between  antiviral drug  and virus load }
 
In the last few decades, mathematical modeling  along with  antiviral drugs   helps  to develop  a therapeutic strategy.   The first model that describes the dynamics of  host cells $x$,  virus load $v$, and  infected cells $y$ as a  function of time $t$  was  introduced in the works \cite{mu1,mu2, mu3, mu4, mu5}. Accordingly, the dynamics of  the host cell, infected cell, and  virus is  governed  by  
\begin{eqnarray}
{\dot x}&=&\lambda-dx-\beta x v, \nonumber \\
{\dot y}&=&\beta x v-ay,  \nonumber \\
{\dot v}&=&ky-uv.
\end{eqnarray}
The   host cells  are produced at rate of  $\lambda$  and die naturally  at a constant rate $d$ with a half-life of  $x_{t_{1\over 2}}={ln(2)\over d}$. The  target cells become infected at a rate of $\beta$ and die at a rate of $a$  with a corresponding half-life of  $y_{t_{1\over 2}}={ln(2)\over a}$. On the other hand,   the  viruses   reproduce at a rate of $k$ and die with a rate  $u$ with a half-life of  $v_{t_{1\over 2}}={ln(2)\over u}$ \cite {mu10,mu11,mu12,mu14}. In this model, only  the interaction between  the host cells and   viruses  is considered neglecting  other cellular activities. The host cells, the infected cells, and the  viruses   have  a lifespan of $1/d$, $1/a$, and $1/u$, respectively.  During the  lifespan of  a cell, one infected cell produces $N=u/a$  viruses on average \cite{muu2, mu12,mu14}.

The capacity  for the virus to spread   can be determined via  the basic reproductive ratio  
\begin{eqnarray}
R_{0}={\lambda \beta k \over a d u}.
\end{eqnarray}
Whenever  $R_{0}>1$, the virus spreads  while 
when $R_{0}<1$  the virus will be  cleared by the host immune system \cite{muu2}.

To examine   the dependence  of  uninfected cells, infected cells and virus load on the system  parameters during antiviral treatment, the above  model system (Eq. (1))  was  modified  by Neumann $et.$ $al.$ \cite{mu8}  and Dahari $et.$ $al.$ \cite{mu9}. The modified  mathematical model presented in those  works analyzes the  effect of antiviral  drugs that either prevents infection  of new cells ($e_{k}$) or  stops  production of  the virion ($e_{p})$.  In this case, the above equation can be remodified to include  the effect of antiviral drugs  as 
\begin{eqnarray}
{\dot x}&=&\lambda-dx-(1- e_k)\beta x v \nonumber \\
{\dot y}&=&(1-e_k)\beta x v-ay  \nonumber \\
{\dot v}&=&(1-e_p)ky-uv
\end{eqnarray}
where the terms  $e_k$  and $e_p$ are  used  when the antivirus blocks infection and  virion production, respectively. For instance, $e_p=0.8$ indicates  that  the drug has  efficacy in blocking virus  production by  $80$ percent.  The antiviral drug  such as protease inhibitor  inhibits  the infected cell  from producing  the right gag protein  as a result the virus becomes noninfectious. A drug such as a reverse transcriptase inhibitor  prohibits    the infection of new cells.

Moreover, the results  obtained  in the last few decades  depict  that,  usually HIV patient  shows high viral load  in the first few weeks of  infection. As a result, the viral load  becomes  the highest  then it starts declining  for a few weeks.  The viruses then keep replicating  for  many years   until the patient  develops AIDS.  Since  virus replication is prone to errors, the virus often develops  drug resistance; HIV  mutates to  become drug-resistant.  Particularly  when the antiviral drug is administered  individually, the ability of the  virus to develop   drug resistance steps up.  However, a triple-drug therapy which includes one protease inhibitor  combined with  two reverse transcriptase  inhibitors helps  to reduce the viral load for many years \cite{muu2}.

Since the antiviral drugs are sensitive to time,  to include this effect, 
next,  we will  modify the Neumann  and Dahari  model.

{\it Case one.\textemdash}As discussed before, the effect  of antiviral  drugs is short-lived since the virus  mutates  once the drug is administrated. To include this effect, we  modify the above  equation  by assuming that  $e_P=e_{p}'e^{-rt}$  and  $e_r=e_{r}'e^{-rt}$. The efficacy of  the drugs declines exponentially  as time progresses.  The decaying  rate aggravates when $r$ tends to increase.  Hence  let   us rewrite Eq. (3) as 
 \begin{eqnarray}
{\dot x}&=&\lambda-dx-(1- e_{k}'e^{-rt})\beta x v \nonumber \\
{\dot y}&=&(1-e_{k}'e^{-rt})\beta x v-ay  \nonumber \\
{\dot v}&=&(1-e_{p}'e^{-rt})ky-uv.
\end{eqnarray}

The term related to the reproductive ratio  is given as 
\begin{eqnarray}
R_{1}={\lambda \beta k \over a d u} (1-e_{k}')(1-e_{p}').
\end{eqnarray}
When  $R_{1}<1$,  the antivirus drug   is capable of  clearing  the virus    and  if $R_{1}>1$, the  virus tends to spread. 
At steady state, 
 \begin{eqnarray}
\overline{x}&=&{au\over \beta k} \nonumber \\
\overline{y}&=&{\beta \lambda k-a d u \over a \beta  k}  \nonumber \\
\overline{v}&=&{\beta \lambda k-a d u \over a \beta  u}.
\end{eqnarray}
As one can note that,   when $t\to \infty$, $e_{p} \to 0$ and $e_{k} \to 0$. At  steady state, only one newly  infected cell  arises   from one infected cell \cite{muu2}.

\begin{figure}[ht]
\centering
{
    \includegraphics[width=6cm]{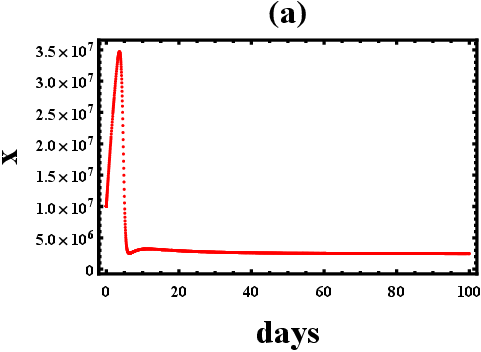}}
\hspace{1cm}
{
    \includegraphics[width=6cm]{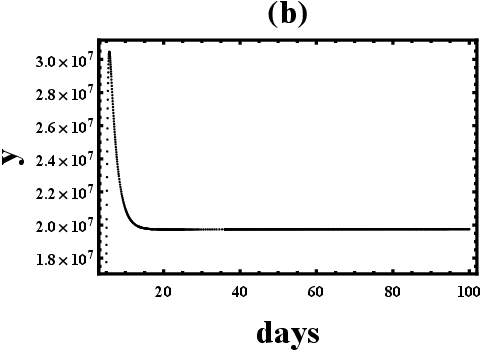}
}
{
    \includegraphics[width=6cm]{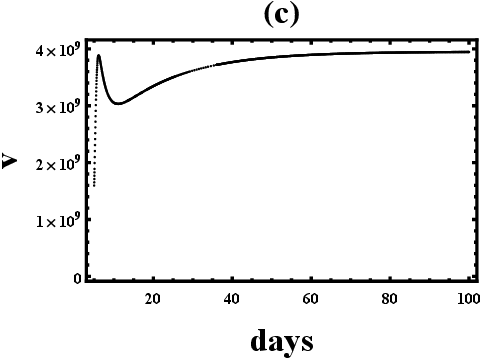}
}
\caption{ (Color online) (a) The number of host cells $x$ as  a function of time (days). (b) The number of infected cells as function of time  (days). (c) The  virus load as  a function of the  time (days). In the figure, we fix  $\lambda  = 10^7$,  $d = 0.1$, $a = 0.5$, $\beta = 2X10^{-9}$, $k = 1000.0$,  $u = 5.0$, $e_p = 0.5$, $e_k = 0.5$ and $r=0.06$. } 
\label{fig:sub} 
\end{figure}
Let us next explore how the number of host cells $x$, the  number of infected cells $y$, and  the viral load $v$ behave as a function of time by exploiting Eq. (4) numerically. From now on, all of the physiological parameters  are considered to vary per unit time (days). Figure 1 depicts the  plot of  the number of host cells $x$, the number of infected cells $y$ and  the number of virus  as function of  time (days) for parameter choice of $\lambda  = 10^7$,  $d = 0.1$, $a = 0.5$, $\beta = 2X10^{-9}$, $k = 1000.0$,  $u = 5.0$, $e_p= 0.5$, $e_k= 0.5$ and $r=0.06$. The  figure depicts that  in the presence of an antiviral  drug, the number of  $CD_{4}$  cells increases and attains a maximum value. The cell numbers  then  decrease and  saturate to a constant value. The number of  infected cells decreases and saturates to a constant value. On the other hand,  the viral load decreases as the antiviral takes an effect. However, this effect is short-lived since  the  
 the  viral load   increases back  as the viruses  develop   drug resistance.

\begin{figure}[ht]
\centering
{
    \includegraphics[width=6cm]{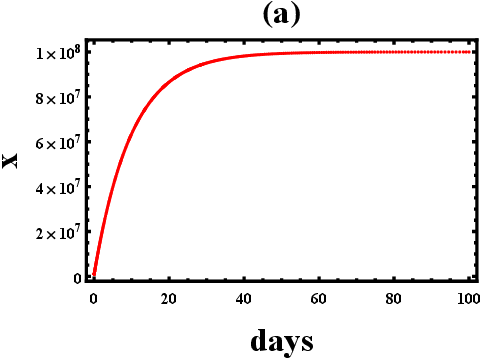}}
\hspace{1cm}
{
    \includegraphics[width=6cm]{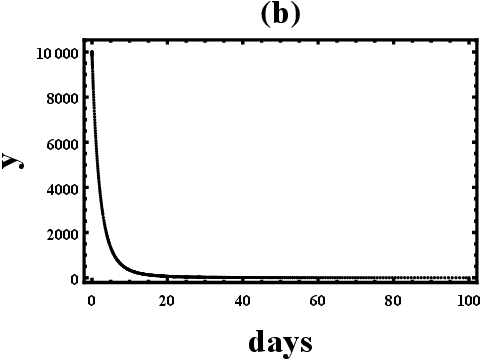}
}
{
    \includegraphics[width=6cm]{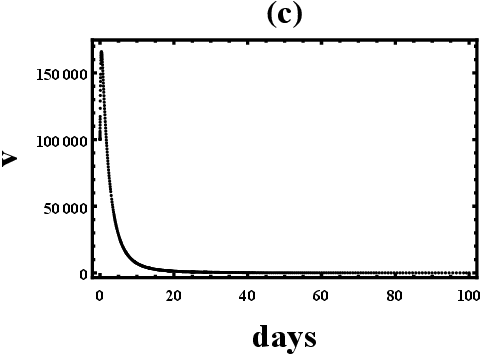}
}
\caption{ (Color online) (a) The number of host cells $x$ as  a function of time (days). (b) The number of infected cells as function of time (days). (c) The  plasma virus load as  a function of time (days). In the figure, we fix  $\lambda  = 10^7$,  $d = 0.1$, $a = 0.5$, $\beta = 2X10^{-9}$, $k = 1000.0$,  $u = 5.0$, $e_p = 0.9$, $e_k = 0.9$ and $r=0.0001$. } 
\label{fig:sub} 
\end{figure}

When $r$  is small,   the ability of  the  antiviral drug  to overcome drug  resistance increases. As depicted in Fig. (2),  for very small $r$, the host cells  increase in time, and at a steady state, the cells  saturate to a constant value.  On the contrary, the infected cells as well as the plasma virus load monotonously decrease as  time progresses. The figure is plotted by fixing    $\lambda  = 10^7$,  $d = 0.1$, $a = 0.5$, $\beta = 2X10^{-9}$, $k = 1000$,  $u = 5.0$, $e_p= 0.9$, $e_k= 0.9$ and $r=0.0001$. These all results  indicate that when combined drugs are administrated,  the viral load  is significantly  reduced  depending on the initial viral load.  
\begin{figure}[ht]
\centering
{
    \includegraphics[width=6cm]{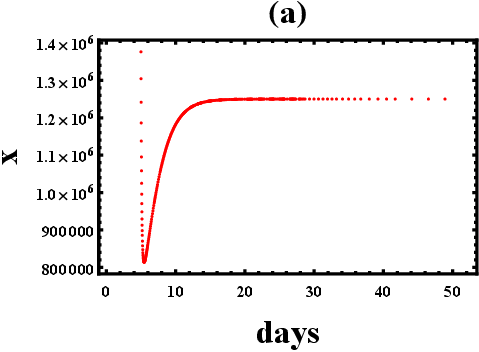}}
\hspace{1cm}
{
    \includegraphics[width=6cm]{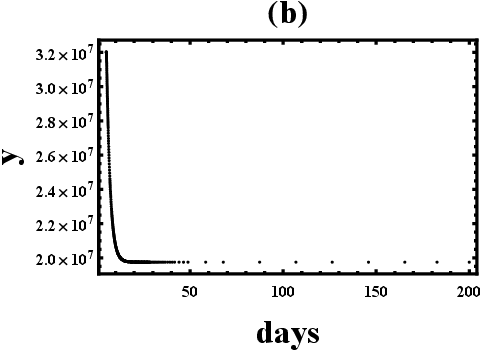}
}
{
    \includegraphics[width=6cm]{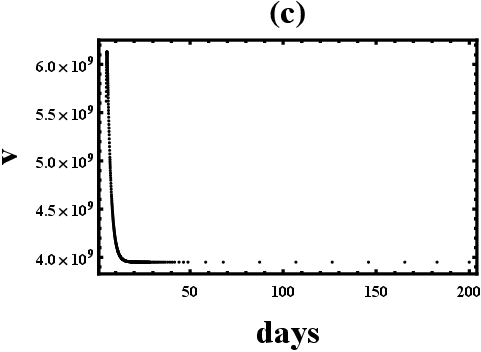}
}
\caption{ (Color online) (a) The number of host cells $x$ as  a function of days. (b) The number of infected cells as function of days. (c) The  virus load as  a function of the  days. In the figure, we fix  $\lambda  = 10^7$,  $d = 0.1$, $a = 0.5$, $\beta = 2X10^{-9}$, $k = 1000.0$,  $u = 5.0$, $e_p = 0.9$, $e_k = 0.9$ and $r=20.0$. } 
\label{fig:sub} 
\end{figure}
Figure 3 is plotted by fixing 
$\lambda  = 10^7$,  $d = 0.1$, $a = 0.5$, $\beta = 2X10^{-9}$, $k = 1000$,  $u = 5.0$, $e_p = 0.9$, $e_k = 0.9$ and $r=20.0$. The figure exhibits that   for large $r$, the $CD_{4}$  cells  decrease  and exhibit a local minima. As time progresses, the number of cells   increases and saturates to a constant value. On the other hand, the number of  infected  cells $y$ and  the viral load $v$  decreases  and saturates to considerably  large    value as time progresses.  Figure 3 also depicts that when the drug is unable to control the infection in a very short period of time, the number of drug resistant viruses steps up.

{\it Case two.\textemdash} In the previous case, the  efficacy  of the drug is considered  to   decrease  monotonously   as time progresses.  In this section,   the  efficacy  of the  drug is assumed  to  decrease to a constant value as time  increases  maintaining  the relation   $e_P=e_{p}'(1+e^{-rt})/m$  and  $e_r=e_{r}'(1+e^{-rt})/m$. 
 The dynamics of 
host cells, infected cells, and viral load  is governed by the equation 
\begin{eqnarray}
{\dot x}&=&\lambda-dx-(1- e_{r}'(1+e^{-rt})/m)\beta x v \nonumber \\
{\dot y}&=&(1-e_{r}'(1+e^{-rt})/m)\beta x v-ay  \nonumber \\
{\dot v}&=&(1-e_{p}'(1+e^{-rt})/m)ky-uv.
\end{eqnarray}
After some algebra, the term related to the  basic reproductive ratio  reduces to 
\begin{eqnarray}
R_{1}={\lambda \beta k \over a d um^2} \left(m-2e_{k}'\right)\left(m-2e_{p}'\right).
\end{eqnarray}
As one can see  from Eq. (8) that as $m$ steps up, the drug losses its potency  and as a result  $R_{1}$ increases.   When $R_{1}<1$,  the antivirus drug   treatment is successful  and  this occurs  for large  values of   $e_p $ and  $e_k $.  When $R_{1}>1$, the  virus overcomes the antivirus  treatment.

 At equilibrium, one finds 
 \begin{eqnarray}
\overline{x}&=&{au m^2\over \beta k (m-e_r')(m-e_p')} \nonumber \\
&=& { \lambda  \over d R_{0} }{ m^2\over  (m-e_r')(m-e_p')}\nonumber \\
\overline{y}&=&{ \lambda  \over a }+ {du m^2\over \beta k (e_r'-m)(m-e_p')} \nonumber \\
&=& \left(R_{0}-{ m^2\over  (m-e_r')(m-e_p')} \right){du\over \beta k} \nonumber \\
\overline{v}&=&{dm\over (\beta e_k'-\beta m})+{\lambda k(m-e_p')\over (amu)} \nonumber \\
&=&\left({R_{0}(m-e-p')\over m}-{m\over (m-e_k')}\right){d\over \beta}.
\end{eqnarray}
The case  $R_{0} \gg 1$ indicates that  the equilibrium abundance of  the uninfected cells  is much less  than the  number of uninfected  cells  before treatment. When the drug is successful, (large values  of  $e_p'$ or $e_r'$),   the equilibrium abundance of the uninfected cells increases. On the contrary, for a highly cytopathic virus ( $R_{1}\gg 1$ ),   the number of infected cells, as well as   the viral load   steps up. When $e_p'$ and $e_r'$  increase,  the  equilibrium abundance of  infected cells as well as viral  load  decreases.

In general for large $R_{0}$,  Eq. (9) converges to  
\begin{eqnarray}
\overline{y}&=&{ \lambda   \over a }\nonumber \\
\overline{v}&=&{\lambda k(m-e_p')\over (amu)}.
\end{eqnarray}
Clearly $\overline{v}$ decreases as  $e_p'$ and $e_r'$  increase.

The overall efficacy  can be written as  \cite{mu9}
\begin{eqnarray}
1-e=\left(1-{e_{r}'\left(1+e^{-rt}\right)\over m}\right)\left(1-{e_{p}'\left(1+e^{-rt}\right)\over m}\right)
\end{eqnarray}
where $0< e_{r}'<1$  and $0<e_{p}'<1$.
At steady state  $1-e=\left(1-{e_{r}'\over m}\right)\left(1-{e_{p}'\over m}\right)$.  The transcritical bifurcation point  (at steady state) can be analyzed  via Eq. (9) and after some algebra we find 
\begin{eqnarray}
1-e=\left(1-{e_{r}'\over m}\right)\left(1-{e_{p}'\over m}\right)={adum \over \lambda \beta k}={x_{1}\over x_{0}}
\end{eqnarray}
where $x_{0}={\lambda \over d}$ denotes the number of uninfected host cells before infection  and $x_{1}={aum \over \beta k}$ designates the number of uninfected cells in the chronic case.  This implies the critical efficacy is given as $e_{c}=1-{x_{1}\over x_{0}}=1-{adum \over \lambda \beta k}$.

To   write the  overall efficacy as a  function of time,  for simplicity, let us further assume that $e_{r}'=e_{p}'$. In this case,  Eq. (12) can be rewritten as 
\begin{eqnarray}
{e_{r}'\over m}=1\pm \sqrt{{adum \over \lambda \beta k}}
\end{eqnarray}
and hence 
\begin{eqnarray}
1-e= \left(1-\left(1\pm \sqrt{{adum \over \lambda \beta k}}\right)\left(1+e^{-rt}\right)\right)^2.
\end{eqnarray}
From Eq. (14), one finds
\begin{eqnarray}
e_c= 1-\left(1-\left(1\pm \sqrt{{adum \over \lambda \beta k}}\right)\left(1+e^{-rt}\right)\right)^2.
\end{eqnarray}
The  critical efficacy  serves as an  alternative  way of  determining  whether  antiviral treatment is successful or not.  When $e>e_c$, the antiviral  clears the infection and if  $e<e_c$, the virus replicates.

\section{ The correlation between  antiviral drug,  immune response  and virus load }

The basic mathematical model  that  specifies   the relation between  the  immune response, antiviral drug, and  viral load   is given by 
\begin{eqnarray}
{\dot x}&=&\lambda-dx-(1- e_{r}'(1+e^{-rt})/m)\beta x v \nonumber \\
{\dot y}&=&(1-e_{r}'(1+e^{-rt})/m)\beta x v-ay-pyz  \nonumber \\
{\dot v}&=&(1-e_{p}'(1+e^{-rt})/m)ky-uv  \nonumber \\
{\dot z}&=&c-bz.
\end{eqnarray}
Once again  the terms $x$, $y$,  and $v$ denote  the uninfected cells, infected cells, and the viral load. 
The term  $z$  denotes the CTL  response and the CTL  die at a  rate of $b$ and produced at a rate of $c$.  The term CTL  is defined as cytotoxic lymphocytes   that  have  responsibility  for killing the  infected cells. 
The term related to the basic reproduction rate is given as   
\begin{eqnarray}
R_{1}={\lambda \beta k \over (a+{cp\over b}) d um^2} (m-2e_{k}')(m-2e_{p}').
\end{eqnarray}
It vital to see that when $R_{0}>0$  the virus  becomes successful   to persist an infection which triggers an immune responses $z$. As long as  the coordination between   the immune response  and   the antiviral  drug treatment   is  strong enough, the virus will be cleared  $R_{1}<1$. As it can be clearly seen  from Eq. (17), when the CTL response step up, $R_{1}$ declines as expected.

 The equilibrium abundance of the host cells, infected cells, viral load,  and CTL  response  can be given as
 \begin{eqnarray}
\overline{x}&=&{m^2 u( ab+cp)\over \beta b k  (m-e_{k}')(m-e_{p}')} \nonumber \\
&=& { \lambda  \over a d R_{0} }{ ( ab+cp)m^2\over  (m-e_r')(m-e_p')}\nonumber \\
\overline{y}&=&{ b \lambda  \over (ab +cp) }+ {du m^2\over \beta k (e_r-m)(m-e_p)} \nonumber \\
&=& \left({abR_{0}\over (ab +cp)}-{ m^2\over  (m-e_r')(m-e_p')} \right){du\over \beta k}\nonumber \\
\overline{v}&=&{dm\over (\beta e_k-\beta m})+{b \lambda k(m-e_p)\over mu(ab +cp)} \nonumber \\
&=&\left({bR_{0}(m-e-p')\over m (ab+cd)}-{m\over (m-e_k')}\right){d\over \beta} \nonumber \\
\overline{z}&=&{c\over b}.
\end{eqnarray}
Exploiting Eq. (18), one can deduce when  $R_{0}\gg 1$, the  equilibrium abundance of the uninfected cells  becomes much lower  in comparison to the number of cells before treatment.   The equilibrium  abundance of the uninfected cells steps up when there is  CTL response (when $c$ and $p$  increase) or when the  antiviral   treatment is successful (when $e_p'$ and $e_r'$  increase). On the contrary,  $\overline{y}$  and $\overline{v}$ decline  whenever there is a strong CTL response  or  when the antiviral treatment is successful.  

In Fig. 5a , we plot the phase diagram   for a regime    $R_{1}<1$ (shaded region) in the phase space of $e_k$ and $e_p$. In Figure 5b, the phase diagram for  a regime  $R_{1}<1$ (shaded region)  in the phase space of $m$  and $e_p=e_k$ is plotted. In the figure, we fix  $\lambda  = 10^7$,  $d = 0.1$, $a = 0.5$, $\beta = 2  X10^{-9}$, $k = 1000$,  $u = 5$, $m=2$,  $r=0.0001$, $p=1.0$, $b=0.5$ and $c=2.0$.

Furthermore,  the transcritical bifurcation point can be analyzed  via Eq. (18) and after some algebra (at steady state)  we find 
\begin{eqnarray}
1-e=\left(1-e_{r}'/m \right )\left(1-e_{p}'/m\right)={(ab+cp ) dum \over \lambda \beta k b}={x_{1}\over x_{0}}
\end{eqnarray}
where $x_{0}={\lambda \over d}$ denotes the number of uninfected host cells before infection  and $x_{1}={(ab+cp)um \over \beta k b}$ designates the number of uninfected cells in a chronic case.  This implies the critical efficacy is given as $e_{c}=1-{x_{1}\over x_{0}}=1-{(ab+cp)um d\over \beta k b \lambda}$.

\begin{figure}[ht]
\centering
{
    \includegraphics[width=6cm]{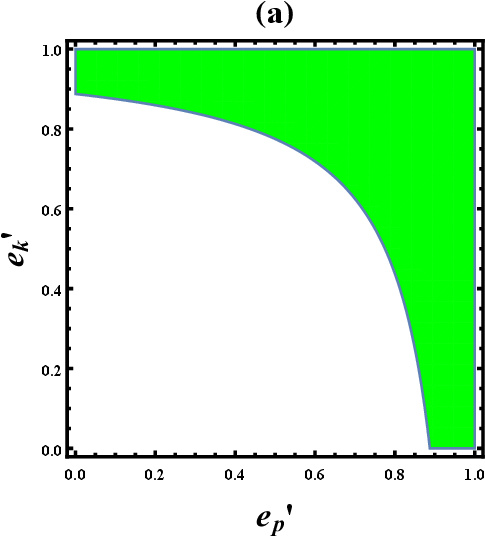}}
\hspace{1cm}
{
    \includegraphics[width=6cm]{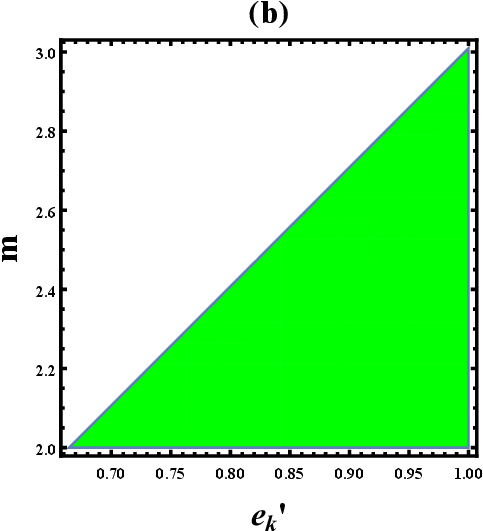}
}
\caption{ (Color online) (a) The phase diagram for a regime  $R_{0}<1$  in the phase space of $e_k$ and $e_p$. (b) The phase diagram for a regime   $R_{0}<1$  in the phase space of $m$  and $e_p=e_k$. In the figure, we fix  $\lambda  = 10^7$,  $d = 0.1$, $a = 0.5$, $\beta = 2  X10^{-9}$, $k = 1000$,  $u = 5$, $m=2$,  $r=0.0001$, $p=1.0$, $b=0.5$ and $c=2.0$. } 
\label{fig:sub} 
\end{figure}
Assuming  $e_{r}'=e_{p}'$,  Eq. (19) can be rewritten as 
\begin{eqnarray}
e_{r}'/m=1\pm \sqrt{{(ab+cp)dum \over \lambda \beta k b}}
\end{eqnarray}
and the  overall efficacy as a  function of time is given by 
\begin{eqnarray}
\left(1-e \right)= \left(1-\left(1\pm \sqrt{{(ab+cp)dum \over \lambda \beta k b}}\right)\left(1+e^{-rt}\right)\right)^2.
\end{eqnarray}
From Eq. (21), one finds
\begin{eqnarray}
e_c= 1-\left(1-\left(1\pm \sqrt{{\left(ab+cp \right)dum \over \lambda \beta k b}}\right )\left(1+e^{-rt}\right)\right)^2.
\end{eqnarray}
Once again,  the infection will be cleared 
when $e>e_c$, and the virus replicates as long as  $e<e_c$.

\begin{figure}[ht]
\centering
{
    \includegraphics[width=6cm]{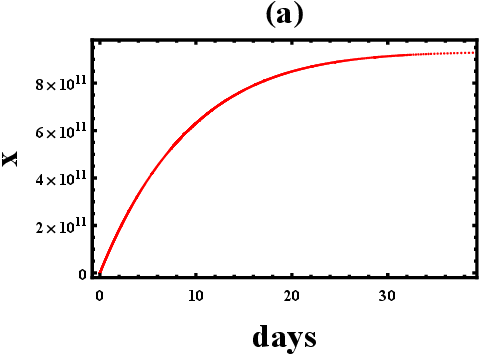}}
\hspace{1cm}
{
    \includegraphics[width=6cm]{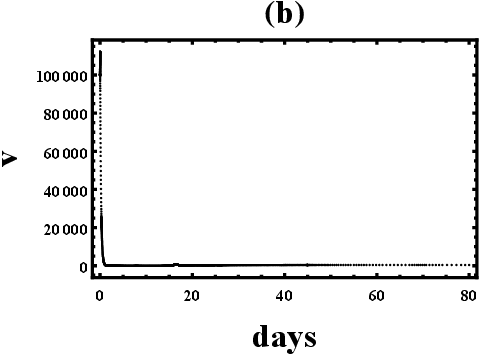}
}
{
    \includegraphics[width=6cm]{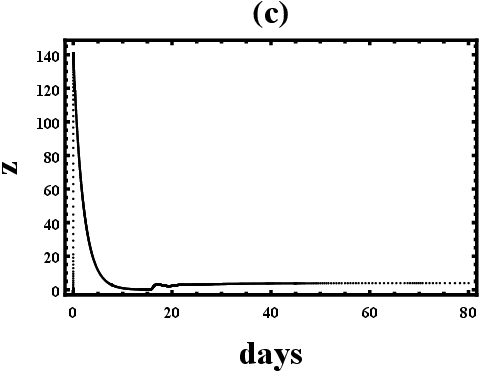}
}
\caption{ (Color online) (a) The number of host cells $x$ as  a function of days. (b) The number of infected cells as function of days (c) The  virus load as  a function of the  days. In the figure, we fix  $\lambda  = 10^7$,  $d = 0.1$, $a = 0.5$, $\beta = 2X10^{-9}$, $k = 1000$,  $u = 5.0$, $e_p' = 0.9$, $e_k' = 0.9$, $r=0.1$, $p=10.0$, $b=0.5$ and $c=1.0$. } 
\label{fig:sub} 
\end{figure}
In order to get a clear insight, let us   explore Eq.  (16). In Fig. 6, the number of host cells $x$, the number of infected cells and the  viral  load as  a function of time is plotted. In the figure, we fix  $\lambda  = 10^7$,  $d = 0.1$, $a = 0.5$, $\beta = 2X10^{-9}$, $k = 1000$,  $u = 5.0$, $e_p' = 0.9$, $e_k' = 0.9$, $r=0.1$, $p=10.0$, $b=0.5$ and $c=1.0$.
For such parameter choice, $R_{0} \gg 1.0$  and $R_{1} \ll 1.0$  revealing that initially, the   virus establishes an infection  but latter the antiviral drug and 
 CTL response  collaborate  to clear the infection. As a result, the  number of host cells increases while the viral load as well as the infected cells decreases. Since the  virus is responsible  for initiating  the   CTL response, as the viral load declines, the  CTL response  step down.

\section{The dynamics of mutant virus in the presence of antiviral drugs }
As discussed before,  the fact that  viruses are an obligate parasite of the cells forces drug  discovery  to be  complicated as  the  drug’s adverse effects directly affect the host cells.  Many medically important   viruses  are also   virulent and hence they cannot be propagated or tested via animal models.  This in turn   forces  the drug discovery to be  lengthy.    Moreover, unlike  other antimicrobial drugs, antiviral drugs  have to be 100 percent potent to completely avoid drug resistance.   In other words, if the drug partially inhibits the replication of the virus,    through time, the   number of  the resistant virus  will dominate the  cell culture.  

To discuss the dynamics of the mutant virus,  let us assume  that 
  the virus mutates (when a single drug is administered )  by  changing  one base every $10^{4}$  viruses. If $10^{11}$  viruses are produced per day, then this results in $10^{7}$ mutant  viruses. On the contrary,  when two antiviral drugs are administrated,  $10^{3}$ mutant viruses will be produced. In the case of triple drug therapy, no mutant virus is produced.  To account for  this effect,  let  us remodify the rate of mutant virus production per day  as
\begin{eqnarray}
k= 10^{11-4s}, {s=1,2,3}
\end{eqnarray}
where the variable $s=1,~2,~ 3$ corresponds to  a single, double, and triple drug therapy, respectively.  
    
Here  we assume only the rate of mutant virus production $k$ determines  the dynamics and  $10^{11}$ viruses are produced per day.
To get an instructive   analytical solution regarding   the relation   between antiviral drug and viral load, let us solve the differential  equation 
\begin{eqnarray}
{\dot v}&=&k-uv
\end{eqnarray}
neglecting  the effect of  uninfected and infected host cells.
Here the mutant virus  produced at  rate  of $k$ and  die   with the rate of $u$.     The solution   for the above equation is given as 
\begin{eqnarray}
v(t)={e^{-ut}(-k+k e^{ut}+uN )\over u}.
\end{eqnarray}
Whenever ${k \over u}>1$ the virus spreads and when ${k\over u}<1$, the antiviral  is capable of eliminating the virus.

 Exploiting Eq. (25) one  can comprehend that, in the case of  single therapy, the virus load decreases during the course of treatment. As time progresses, the viral load increases back due to the emergence of  drug resistance (see Fig. 8a). In the case of double drug therapy, as shown in Fig. 8b, the viral load decreases   but relapses back as time progresses.  When triple  drugs are administered, the   viral replication becomes suppressed as depicted in Fig 8c.  The readers should understand that the triple drug therapy does not guarantee a cure. If the patient  halts  his or her therapy,  the viral replications will resume  because of the latent and chronic   infected cells.  
At steady state,  we get 
\begin{eqnarray}
\overline{v}={k\over u}
\end{eqnarray}
At equilibrium, the  viral load spikes as $k$ increases and it decreases as $u$ steps up.
\begin{figure}[ht]
\centering
{
    \includegraphics[width=6cm]{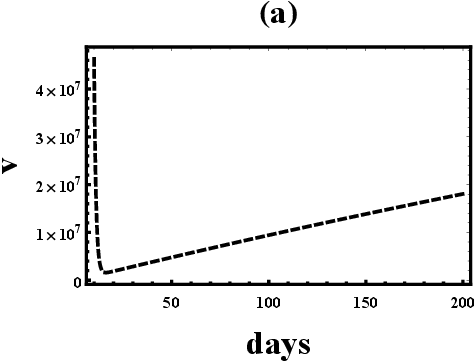}}
\hspace{1cm}
{
    \includegraphics[width=6cm]{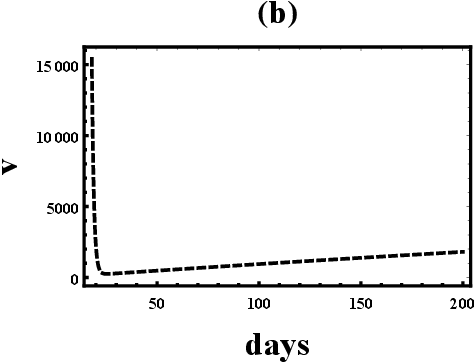}
}
{
    \includegraphics[width=6cm]{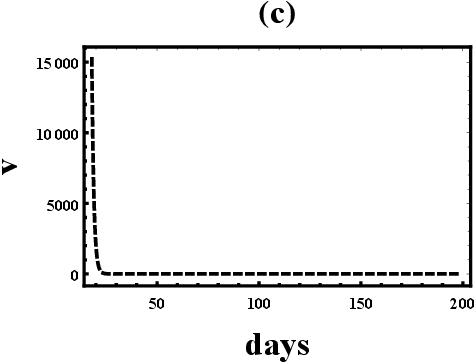}
}
\caption{ (Color online) The  virus load as  a function of time. In the figure, we fix   $u = 1.0$.
 (a) Single drug therapy $s=1$. (b) Double  drug therapy $s=2$.  (c) Triple drug therapy $s=3$.  } 
\label{fig:sub} 
\end{figure}

\section{Summary and conclusion}

Developing  antiviral   drugs is challenging  but   an urgent  task  since 
outbreaks of viral diseases not only killed several people but also cost trillion dollars worldwide.  The discovery of new antiviral  drugs together with emerging mathematical models helps to  understand the dynamics of the virus in vivo.  For instance, the pioneering  mathematical models on HIV   shown  in the works  \cite{mu1,mu2,mu3,mu4,mu5,mu6,mu7} disclose   the host-virus  correlation. Moreover,  
to study  the correlation between, antiviral drugs and  viral load, an elegant  mathematical model was   presented in the works  \cite{mu8, mu9}.

Due to the  emergence of drug resistance,   the efficiency   of antiviral  drugs is short-lived. To study  this effect, we  numerically study  the dynamics  of the host cells and viral load  in the presence of an  antiviral  drug that either prevents infection ($e_{k}$) or  stops the production of virus ($e_{p}$).  For  the drug whose efficacy depends on time,  we show that     when the efficacy  of the  drug is low, the viral load decreases  and  increases back  in time    revealing the effect of the antiviral  drugs is  short-lived.  On the contrary, for the antiviral drug  with high efficacy, the viral load, as well as the number of infected cells, monotonously   decreases while  the number of uninfected cells   increases. The dynamics  of   critical drug  efficacy   on time is also  explored. Furthermore,  the correlation between  viral load, an antiviral drug,  and  CTL  response  is  also explored. Not only the  dependence for the basic  reproduction ratio on the  model parameters is   explored but also we   analyze     the critical drug  efficacy   as a function of time.  The term related to the  basic reproduction ratio  increases when  the CTL  response  step up.    A simple analytically solvable  mathematical model  to analyze the correlation between  viral load  and antiviral  drugs is also presented.

In conclusion, in this work,  we present  a  simple model which not only   serves as a basic tool  for   better understanding   of  viral dynamics in vivo and vitro but also helps in   developing an effective therapeutic strategy.

\section*{Acknowledgment}
I would like to thank Mulu  Zebene  and Blaynesh Bezabih for the constant encouragement.

\end{document}